\documentclass{an}
\usepackage{graphicx}
\usepackage{times}
\usepackage{fancyhdr}
\begin{document}

\title{The formation of brown dwarfs}

\author{A.P. Whitworth\inst{1} \and S.P. Goodwin\inst{1}}
\institute{School of Physics \& Astronomy, Cardiff University, 5 The Parade,
Cardiff CF24 3YB, Wales, UK}

\date{Received; accepted; published online}

\abstract{We review four mechanisms for forming brown dwarfs: (i) turbulent 
fragmentation (producing very low-mass prestellar cores); (ii) gravitational 
instabilities in discs; (iii) dynamical ejection of stellar embryos from their 
placental cores; and (iv) photo-erosion of pre-existing cores in HII regions. 
We argue (a) that these are simply the mechanisms of {\it low-mass star 
formation}, and (b) that they are not mutually exclusive. If, as seems 
possible, all four mechanisms operate in nature, their relative importance 
may eventually be constrained by their ability to reproduce the binary 
statistics of brown dwarfs, but this will require fully 3-D radiative 
magneto-hydrodynamic simulations.
\keywords{Star formation - brown dwarfs}}

\correspondence{ant@astro.cf.ac.uk}

\maketitle

\section{Introduction}

The existence of brown dwarfs was first proposed on theoretical grounds by Kumar 
(1963) and by Hayashi \& Nakano (1963). However, more than three decades then passed 
before brown dwarfs were observed unambiguously (Rebolo et al., 1995; Nakajima et 
al., 1995; Oppenheimer et al. 1995). Brown dwarfs are now observed routinely, and 
are estimated to be comparable in number with hydrogen-burning stars. It is 
therefore appropriate to ask how brown dwarfs form, and in particular to ascertain 
(a) whether brown dwarfs form in the same way as hydrogen-burning stars, and (b) 
whether there is a clear distinction between the mechanisms that produce brown dwarfs 
and those that produce planets.

In Section \ref{SEC:STAR} we argue that brown dwarfs do form in the same way as stars, 
on the grounds that their statistical properties (mass function, binary statistics, 
clustering properties, etc.) appear to form a smooth continuum with those of low-mass 
hydrogen-burning stars. We also suggest that understanding how brown dwarfs form is 
the key to answering a fundamental anthropic question, namely, what determines the 
lower mass limit for star formation, and thereby the likelihood of long-lived stars 
with habitable zones. In Section \ref{SEC:TURB} we consider the formation of brown 
dwarfs by turbulent fragmentation, as suggested by Padoan \& Nordlund (2002), and we 
address the question of whether an isolated 
core of brown-dwarf mass formed in this way can cool sufficiently fast to condense out. 
In Section \ref{SEC:DISC} we consider the formation of brown dwarfs by gravitational 
instabilities in discs. We stress that only in massive discs, and at large radii, can 
fragments of a disc contract and cool sufficiently fast to condense out; closer in they 
are likely to bounce and be shredded. We also point out that, in a dense proto-cluster, 
impulsive interactions between discs, or between a disc and a naked star, should be 
common, and may be necessary to ensure disc fragmentation. In Section 
\ref{SEC:EJEC} we consider the formation of brown dwarfs by the ejection mechanism, as 
suggested by Reipurth \& Clarke (2001). We point out that the requirements for this 
mechanism to operate are very general, and therefore it is likely to occur in nature, 
although it is probably not the only mechanism forming brown dwarfs, given the 
difficulty it has producing close BD-BD binaries. In Section \ref{SEC:EROS} we consider the 
formation of brown dwarfs by photo-erosion of pre-existing cores which are overrun by HII 
regions, as suggested by Hester et al. (1996). We stress that this a very robust mechanism, 
in the sense that it does not require very fine tuning of the parameters; but it is 
also a very inefficient mechanism, in the sense that it requires a very massive inital 
core to form a brown-dwarf, and it clearly cannot deliver the brown dwarfs in 
regions like Taurus. In Section \ref{SEC:CONC} we summarise our review.

\section{Why brown dwarfs appear to form like H-burning stars} \label{SEC:STAR}

We shall assume that brown dwarfs form in the same way as hydrogen-burning stars, i.e. on 
a dynamical timescale, by gravitational instability, and with initially uniform elemental 
composition (reflecting the elemental composition of the interstellar medium out of which 
they form). Thus, by implication, we distinguish brown dwarfs from planets, which form on 
a much longer timescale, by the amalgamation of a rocky core and -- if circumstances allow 
-- the subsequent accretion of a gaseous enevelope, resulting in an initially fractionated 
elemental composition with an overall deficit of volatile/light elements. If this is the 
correct way to view the formation of brown dwarfs, and we argue below that it is, then brown 
dwarfs should not be distinguished from stars; many stars fail to burn helium, and most fail 
to burn carbon, without forfeiting the right to be called stars.

The reason for categorising brown dwarfs as stars is that the statistical properties of brown 
dwarfs appear to form a continuum with those of low-mass hydrogen-burning stars.
 
{\it (i) IMF.} The initial mass function is apparently continuous across the hydrogen-burning 
limit at $\sim 0.075\,{\rm M}_{_\odot}$. This is not surprising, since the processes which 
determine the mass of a star are presumed to occur at relatively low densities and 
temperatures, long before the protostellar material knows whether it will reach sufficiently 
high temperature to burn hydrogen before or after reaching sufficiently high density to be 
supported in perpetuity by electron degeneracy pressure. In the light of this continuity, it 
seems perverse to have to speak of `The IMF for Stars and Brown Dwarfs' when `The Stellar IMF' 
is already quite long enough.

{\it (ii) Clustering properties.} In clusters, brown dwarfs appear to be homogeneously mixed 
with H-burning stars, and their kinematics are also essentially indistinguishable. Although 
they have been searched for -- as possible signatures of formation by ejection -- neither a 
greater velocity dispersion of brown dwarfs in very young clusters, nor a diaspora of brown 
dwarfs around older clusters, has been found.

{\it (iii) Binary statistics.} Here we have to distinguish at least two types of binary system. 

In the first type of binary system, the primary is a Sun-like star and the secondary component 
is a brown dwarf. Amongst this type there is a remarkable lack of close systems (the Brown Dwarf 
Desert). However, at larger separations (semi-major axis $a \stackrel{>}{\sim} 100\,{\rm AU}$), 
brown-dwarf secondaries are quite common. Moreover, the lack of close low-mass secondaries is 
not confined to brown dwarfs. There appears to be a general lack of systems with very low 
mass-ratios, $q \stackrel{<}{\sim} 0.1$.

In the second type of binary system, the primary is a brown dwarf, and therefore the 
secondary is also a brown dwarf (or possibly even a planetary-mass object, if this 
distinction must be made, see below). For brown-dwarf primaries, the multiplicity is  
estimated to be $m \sim 30\,{\rm to}\,40\,\%$, the distribution of semi-major axes 
peaks at $a_{_{\rm PEAK}} \sim 4\,{\rm AU}$ with a logarithmic dispersion 
$\sigma_{_{\log\!10\,a}} \sim 0.6$, and the mean mass-ratio is $\bar{q} \sim 0.7$. 
In comparison, G-dwarf primaries are estimated to have $m \sim 60\,\%$, 
$a_{_{\rm PEAK}} \sim 30\,{\rm AU}$, $\sigma_{_{\log\!10\,a}} \sim 1.6$, and 
$\bar{q} \sim 0.3$. The implication is that, as the primary mass decreases, (i) the 
multiplicity decreases (but only quite slowly), (ii) the distribution of semi-major 
axes shifts to smaller separations and becomes narrower (logarithmically), and (iii) 
the distribution of mass ratios shifts towards unity -- with these trends continuing 
across the divide between brown dwarfs and H-burning stars.

In fact, the situation is even more complicated than this, since there are several 
systems in which (a) the primary is a close binary with Sun-like components (rather 
than a single Sun-like star), and/or (b) the Sun-like primary is orbited at large 
radius ($\stackrel{>}{\sim} 100\,{\rm AU}$) by a close BD-BD binary. However, the 
statistics of these systems are limited.

{\it (iv) Discs, accretion and outflows.} Young brown dwarfs are observed to 
have infrared excesses indicative of circumstellar discs, just like young H-burning 
stars. From their H$\alpha$ 
emission-line profiles, there is also evidence for ongoing accretion onto 
brown dwarfs, and the inferred accretion rates form a continuous distribution with those for 
H-burning stars, fitted approxmately by $\dot{M} \sim 10^{-8}\,{\rm M}_{_\odot}\,{\rm yr}^{-1}\,
\left(M/{\rm M}_{_\odot}\right)^2$. Finally, the spectra of brown dwarfs also show forbidden 
emission lines suggestive of outflows like those from H-burning stars, and recently an 
outflow from a brown 
dwarf has been resolved spatially. Thus, in the details of their circumstellar 
discs, accretion rates and outflows, young brown dwarfs appear to mimic H-burning stars very 
closely, and to differ significantly only in scale.

Given this continuity of statistical properties between brown dwarfs and H-burning stars, 
it is probably unhelpful to distinguish brown dwarfs from stars, and in the rest of the paper
we will only use the H-burning limit at $\sim 0.075 {\rm M}_{_\odot}$ as one of several 
reference points in the range of stellar masses. The D-burning limit at $\sim 0.013\,
{\rm M}_{_\odot}$ therefore falls in the same category. We will then define a star as any 
object forming on a dynamical timescale, by gravitational instability, and therefore with 
uniform interstellar elemental composition. With this definition, there is the distinct 
likelihood of a small overlap between the mass range of stars and that of planets. Given 
that in the immediate future we are unlikely to know too much more than the masses of the 
lowest-mass objects, and certainly not their internal composition, we will simply have to 
accept that there is a grey area in the range $0.001\,{\rm to}\,0.01\,{\rm M}_{_\odot}$ 
which may harbour both stars and planets.

It follows that understanding how brown dwarfs form is important, not just for its own 
sake, but because it is the same as understanding how very low-mass stars form. Thus brown 
dwarf formation is a key part of understanding why most stars have masses in the range 
$0.01\,{\rm M}_{_\odot}$ to $100\,{\rm M}_{_\odot}$, and hence why there are lots of 
hospitable stars like the Sun with long-lived habitable zones, and enough heavy elements 
(C, N, O, Si, Mg, Al, Fe, etc.) to produce rocky planets and life. The 
high-mass cut-off is probably due to the fact that radiation pressure makes it hard to 
form the highest-mass stars; and the low-mass cut-off is probably due to the opacity limit. 
By studying brown dwarf formation we can attempt to confirm and quantify the low-mass cut-off.
 
\section{Formation by turbulent fragmentation} \label{SEC:TURB}

The first possibility that we consider is that the processes forming prestellar 
cores create some prestellar cores with very low masses. Very low-mass cores must 
inevitably spawn very low-mass stars, even if they don't fragment during collapse. 
This is the formation mechanism that has been explored by Padoan \& Nordlund (2002). 
By simulating the development of interstellar turbulence, they show that a wide 
range of dense structures is formed. If those structures which are dense and 
coherent enough to be gravitationally unstable are identified as prestellar cores, 
they have a mass spectrum very similar to the observed stellar IMF. There is support 
for this scheme from the observations of Motte, Andr\'e \& Neri (1998) who show 
that the mass function for cores does indeed appear to echo the stellar IMF. However, 
we note (i) that the core mass function should relate more closely to the system IMF 
(rather than the stellar IMF), and (ii) that the completeness limit of the core mass 
function does not extend to brown-dwarf masses. Moreover, the simulations of Padoan 
\& Nordlund do not include gravity, and they use an isothermal equation of state. 
Therefore they do not address the requirement that dynamically contracting cores 
must be able to radiate away at least half the gravitational potential energy being 
released by condensation.

This requirement is normally referred to as the Opacity Limit (but see Masunaga \& 
Inutsuka, 1999), and is presumed to determine the minimum mass for star formation. 
Traditionally, the Opacity Limit has been evaluated on the basis of the 3-D 
hierarchical fragmentation picture developed by Hoyle (1953). In this picture, a 
large protocluster cloud becomes Jeans unstable and starts to contract. As long as 
the sound speed in the gas remains approximately constant, the increasing density 
reduces the Jeans mass, and so separate parts of the cloud (sub-clouds) become 
Jeans unstable and can contract independently of one another. This process repeats 
itself recursively, breaking the original cloud up into ever smaller and denser 
sub-sub...sub-clouds, until the gas becomes so opaque that it can no longer radiate 
away the gravitational energy being released by contraction. At this stage the gas 
starts to heat up, and fragmentation ceases. This yields a minimum mass in the 
range $M_{_{\rm MIN}} \sim 0.007\,{\rm M}_{_\odot}$ to $0.015\,{\rm M}_{_\odot}$ 
(e.g. Low \& Lynden-Bell, 1976; Rees, 1976; Silk, 1977).

However, it appears that 3-D hierarchical fragmentation does not work. There is no 
evidence for its occuring in nature, nor does it occur in numerical simulations of 
star formation. The reason 3-D hierarchical fragmentation does not work probably 
has to do with the fact that the timescale on which a fragment condenses out in 
3-D is always longer than the timescale on which the parent cloud is contracting. 
Therefore fragmentation, if it occurs at all, is only temporary, and the fragments 
are then merged by the overall contraction of the parent cloud. The only way to avoid 
this is to start with fragments which are widely spaced, but then the rate of accretion 
onto a fragment is very high, and even if it starts off with mass $M_{_{\rm FRAG}} \sim 
M_{_{\rm JEANS}}$, it will be many times more massive by the time its contraction 
becomes non-linear. Thus the values for $M_{_{\rm MIN}}$ quoted in the previous paragraph 
are probably significant underestimates for hierarchical 3-D fragmentation.

It is therefore appropriate to revisit the question of the minimum mass for fragmentation, 
but now using a model which invokes 2-D one-shot fragmentation of a shock-compressed layer. 
We argue that this model is more relevant to the contemporary scenario of `star formation  
in a crossing time' (Elmegreen, 2000), and in particular to the scenario simulated by 
Padoan \& Nordlund (2002). In this scenario star formation occurs in molecular clouds 
wherever two -- or more -- turbulent flows of sufficient density collide with sufficient 
ram pressure to produce a shock-compressed layer out of which prestellar cores can condense. 
The model is `2-D' because fragmentation of a shock-compressed layer is in effect two-dimensional 
(the motions which initially assemble a fragment are largely in the plane of the layer), and 
it is `one-shot' in the sense of not being hierarchical or recursive.

A shock-compressed layer is contained by the ram pressure of the inflowing gas, and until 
it fragments it has a rather flat density profile. If we consider the simplest case of a 
head-on collision between two streams of equal density, the resulting layer fragments at time 
$t_{_{\rm FRAG}}$, whilst it is still accumulating, and the fastest growing fragment has mass 
$m_{_{\rm FRAG}}$, radius $r_{_{\rm FRAG}}$ (in the plane of the layer) and half-thickness 
$z_{_{\rm FRAG}}$ (perpendicular to the plane of the layer) given by
\begin{eqnarray}
t_{_{\rm FRAG}} & \sim & (\sigma / G \rho v)^{1/2} \,, \\
m_{_{\rm FRAG}} & \sim & (\sigma^7 / G^3 \rho v)^{1/2} \,, \\
r_{_{\rm FRAG}} & \sim & (\sigma^3 / G \rho v)^{1/2} \,, \\
z_{_{\rm FRAG}} & \sim & (\sigma^5 / G \rho v^3)^{1/2} \,.
\end{eqnarray}
Here $\sigma$ is the net velocity dispersion in the layer, $\rho$ is the 
pre-shock density in the colliding flows, and $v$ is the relative speed with which the 
flows collide. We note (a) that the fragments are initially flattened objects ($r_{_{\rm FRAG}} 
/ z_{_{\rm FRAG}} \sim v / \sigma \gg 1$); (b) that $m_{_{\rm FRAG}}$ is not simply the 
standard 3-D Jeans mass evaluated at the post-shock density and velocity dispersion -- it is 
larger by a factor $(v / \sigma)^{1/2}$; and (c) that our analysis ignores magnetic fields 
and the possibility that the post-shock gas is turbulent. If present, both  magnetic fields 
and turbulence will act to increase the minimum fragment mass.

\begin{figure*}
\resizebox{\hsize}{!}
{\includegraphics[angle=-90,width=11.0cm]{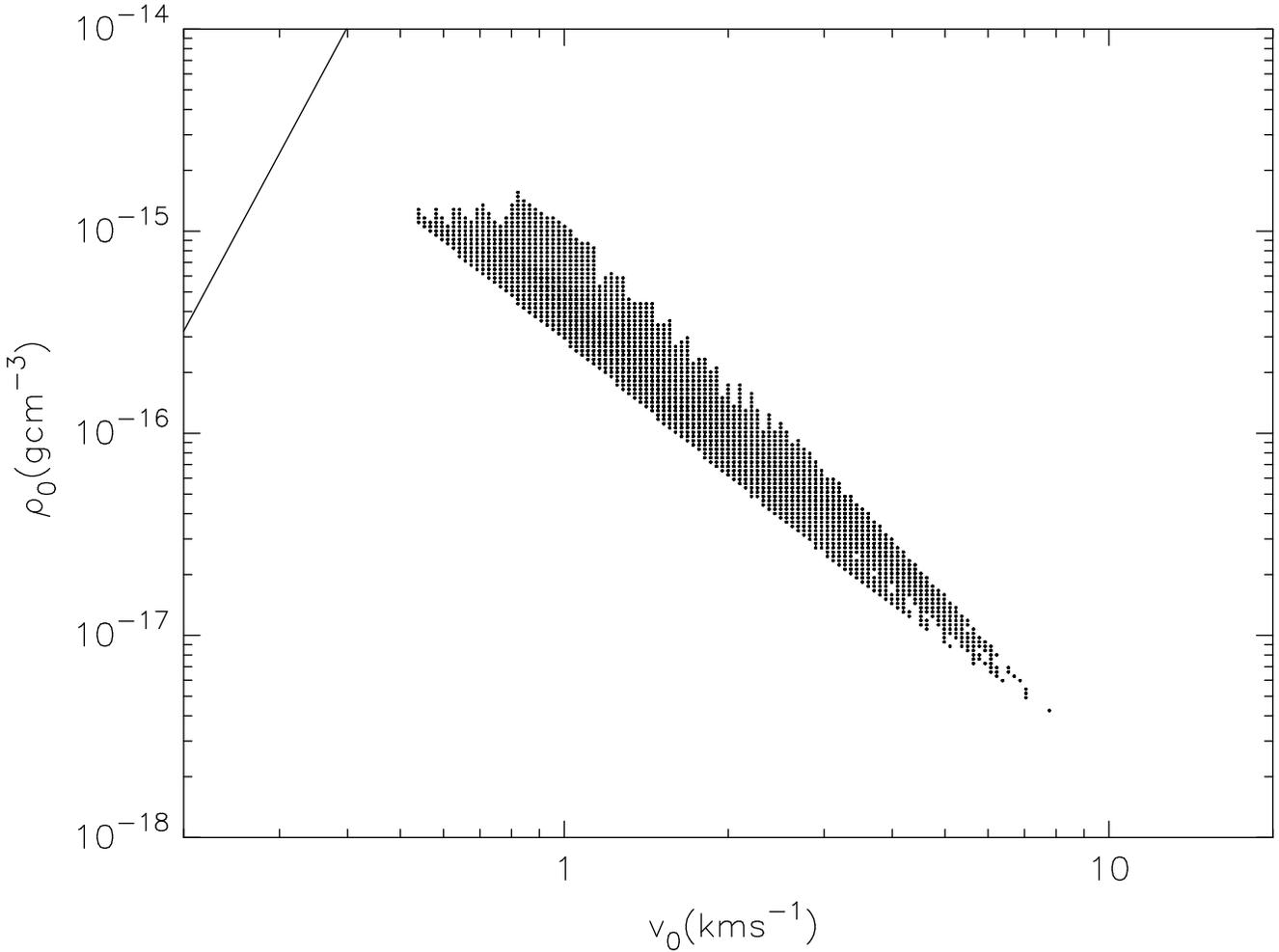}}
\caption{A log/log plot of the $(\rho,v)$ plane. The dots mark 
combinations of pre-shock density. $\rho$, and collision speed, 
$v$, for which the fastest growing fragment has a mass less than 
$0.005\,{\rm M}_\odot$; we assume that the effective post-shock 
sound speed is $\sigma = 0.2\,{\rm km}\,{\rm s}^{-1}$, 
corresponding to molecular gas at $10\,{\rm K}$. The 
irregularities in the boundaries of this region reflect the 
tendency of marginally unstable low-mass fragments to undergo 
pulsations before they collapse. The solid line is the locus 
below which $\rho$ must fall if our treatment of the radiation 
from the accretion shock is to be valid; see Boyd \& Whitworth 
(2005) for details.}
\label{FIG:BOYD}
\end{figure*}

2-D one-shot fragmentation has the advantage that the fastest-condensing fragment has finite 
size, i.e. fragments with initial radius $\sim r_{_{\rm FRAG}}$ condense out faster than 
either larger or smaller fragments. Moreover, we can analyse the growth of a fragment in 
a shock-compressed layer, taking account of the continuing inflow of matter into the 
fragment. Hence we can identify the smallest fragment which can cool radiatively fast 
enough to dispose of {\it both} the $PdV$ work being done by compression, {\it and} the 
energy being dissipated at the accretion shock where matter continues to flow into the 
fragment; these two sources of heat turn out to be comparable. We find (Boyd \& Whitworth, 
2005) that for shocked gas with temperature $T \sim 10\,{\rm K}$ and no turbulence (i.e. 
velocity dispersion, $\sigma$, equal to the isothermal sound speed, 
$0.2\,{\rm km}\,{\rm s}^{-1}$), the smallest fragment which can condense out is less than 
$0.003\,{\rm M}_{_\odot}$, and fragments with mass below $0.005\,{\rm M}_{_\odot}$ 
condense out for a wide range of pre-shock density $\rho$ and shock-speed $v$ (as 
illustrated on Fig. 1). We emphasise that this analysis is more robust than the standard 
one based on 3-D hierarchical fragmentation, on two counts. (i) The fragments have 
condensation timescales shorter than all competing length scales (a well known property 
of layer fragmentation, e.g. Larson, 1985), so they do not tend to merge with their 
neighbours. (ii) Ongoing accretion is taken into account. Indeed, the 
smallest fragment of all starts off with mass $0.0011\,{\rm M}_{_\odot}$, and grows to 
$0.0027\,{\rm M}_{_\odot}$ before its contraction becomes non-linear. We conclude 
that stars with masses down to $0.003\,{\rm M}_{_\odot}$ can condense 
out of shock-compressed layers. If the temperature of the 
post-shock gas can be reduced further still, to below $\sim 6\,{\rm K}$, then it is 
even possible to form `stars' with masses below $0.001\,{\rm M}_{_\odot}$.

\section{Formation by disc fragmentation} \label{SEC:DISC}

Another possibility is that an initially massive prestellar core (i.e. significantly 
more massive than a brown dwarf) spawns brown dwarfs by fragmentation. The fragmentation 
of collapsing cores is a large and complicated topic. However, one of the main 
fragmentation mechanisms which operates in numerical simulations is that a relatively 
massive primary protostar forms, surrounded by a massive disc-like structure (albeit 
not necessarily a relaxed rotationally supported disc), and then lower-mass secondary 
protostars -- including proto--brown-dwarfs -- condense out of the disc (e.g. Bate, 
Bonnell \& Bromm, 2002a,b, 2003; Hennebelle et al., 2004; Goodwin et al. 2004a,b,c). 
This is the mechanism of core fragmentation on which we shall concentrate here.

If we consider a relaxed massive disc in isolation, there is some doubt as to whether it will 
fragment gravitationally, spawning low-mass companions to the central primary protostar, 
or whether spiral modes will act to quickly redistribute angular momentum, thereby 
stabilising -- and ultimately dissipating -- the disc before it can fragment. However, if a 
massive protostellar disc interacts impulsively with another disc, or with a naked star, 
or if the disc simply never has time to relax towards an equilibrium state, then it can 
be launched directly into the non-linear regime of gravitational instability, and 
fragmentation is then much more likely. In the dense proto-cluster environment where 
most protostars are born, such impulsive interactions must be quite frequent. 
Therefore Boffin et al. (1998) and Watkins et al. (1998a,b) have simulated 
parabolic interactions between two protostellar discs, and between a single 
protostellar disc and a naked protostar. All possible mutual 
orientations of spin and orbit are sampled. The critical parameter turns out to be 
the effective shear viscosity in the disc. If the Shakura-Sunyaev parameter is low, 
$\alpha_{_{\rm SS}} \sim 10^{-3}$, most of the secondary protostars have masses in 
the range $0.001\,{\rm M}_{_\odot}$ to $0.01\,{\rm M}_{_\odot}$. Conversely, if 
$\alpha_{_{\rm SS}}$ is larger, $\alpha_{_{\rm SS}} \sim 10^{-2}$, most of the 
secondary protostars have masses in the range $0.01\,{\rm M}_{_\odot}$ to 
$0.1\,{\rm M}_{_\odot}$. The formation of low-mass companions is most efficient 
for interactions in which the orbital and spin angular momenta are all parallel; 
on average 2.4 low-mass companions are formed per interaction in this case. If 
the orbital and spin angular momenta are randomly orientated with respect to each 
other, then on average 1.2 companions are formed per interaction. 

In the above simulations the gas is assumed to behave isothermally, which is 
probably a reasonable assumption, since the discs are large (initial radius 
$1000\,{\rm AU}$) and most of the secondary protostars form at large distance 
(periastra $\stackrel{>}{\sim} 100\,{\rm AU}$). However, disc fragmentation is 
probably not possible at smaller radii because the ambient temperature close 
to the central primary protostar is higher and the surface-density of the disc 
is also higher. Consequently the optical depths through proto-fragments are 
higher and they are unable to cool radiatively sufficiently fast to condense 
out (Rafikov, 2005); instead they contract adiabatically, bounce, and are 
shredded by tidal forces. Thus gravitational fragmentation is probably limited to 
the outer parts of such discs. Rice et al. (2003) present SPH simulations of 
discs fragmentating gravitationally at small radii ($\sim 10\,{\rm AU}$), but 
they use a phenomenological cooling law of the form $du/dt = - 
u/t_{_{\rm COOL}}$, and the values of $t_{_{\rm COOL}}$ which they 
invoke are unrealistically short; also their cooling law seems to admit indefinite 
cooling and their discs appear to fragment only after the cooling becomes 
catastrophic. Boss (2001, 2003) also presents simulations of discs fragmentating 
gravitationally at small radii, performed using a finite difference code with 
radiation transport. However, the reality of the fragments he finds is 
questionable on two counts. First, in evaluating the boundedness of the 
fragments he appears to neglect their internal kinetic energy; 
in a fragment which is bouncing, or contracting but destined to bounce, 
this can be a dominant term in the Virial Theorem. Second, he argues that his 
fragments are cooling by convection, but convection cannot contribute to the 
cooling of a fragment which is condensing out on a dynamical timescale. 
The velocity fields which Boss attributes to convective motions may actually 
be due to bouncing -- in which case they will lead to dissolution of his 
fragments by shredding.

This may help to explain the Brown Dwarf Desert. Brown dwarf companions to 
solar-type primaries can form by disc fragmentation, but only at large radii. 
To end up in closer orbits, they must either accrete material with low 
specific angular momentum, which will tend to increase their mass; or they 
must interact dynamically with a third star, but this tends to place the more 
massive star in the close orbit, and to eject the less massive star (i.e. 
the brown dwarf).

We note that the two mechanisms discussed thus far are not mutually exclusive. 
If the initial prestellar core is already of very low mass, then it will 
inevitably produce a low-mass protostar, irrespective of whether it fragments 
or not. If it has higher mass, it can only produce a very low-mass protostar 
by fragmenting, and one possible mode of fragmention involves the formation 
of a disc.

\section{Formation by ejection} \label{SEC:EJEC}

The collapse of a prestellar core is unlikely to lead to a single star. Even 
quite modest levels of turbulence (e.g. Goodwin, Whitworth \& Ward-Thompson, 
2004a) and/or global rotation (Cha \& Whitworth, 2003; Hennebelle et al., 2004) 
are sufficient to ensure fragmentation. Hence prestellar cores usually spawn 
small-$N$ clusters of protostars ($N \sim 2\,{\rm to}\,6$; e.g. Hubber \& Whitworth, 
2005), which then grow by competitive accretion and interact dynamically 
(Whitworth et al., 1995; Bonnell et al., 2001). Protostars which get ejected 
from the core before they have time to grow to $0.075\,{\rm M}_{_\odot}$ end 
up as brown dwarfs (Reipurth \& Clarke, 2001). It seems inescapable that this 
mechanism occurs in nature, since all that is required is the formation and 
coexistence of more than two protostars in a core, with one of them being less 
massive than $0.075\,{\rm M}_{_\odot}$; $N$-dody dynamics will then almost 
inevitably eject one of the protostars, and usually the least massive one. 

Several numerical simulations have been performed, using SPH with sink particles, 
to demonstrate the viability of this mechanism, both in cores with high levels of 
turbulence (Bate, Bonnell \& Bromm, 2002a,b, 2003; Delgado Donate, Clarke \& Bate, 
2003, 2004), and in cores with low levels of turbulence (Goodwin, Whitworth \& 
Ward-Thompson, 2004a,c). Many of the brown dwarfs formed in these 
simulations retain low-mass discs ($M_{_{\rm DISC}} \stackrel{<}{\sim} 
0.010\,{\rm M}_{_\odot}$ and $R_{_{\rm DISC}} \stackrel{<}{\sim} 40\,{\rm AU}$) 
even after ejection, from which they continue to accrete. They also have a radial 
velocity distribution which is scarcely distinguishable from that of the 
hydrogen-burning stars. This is firstly because part of the overall velocity dispersion 
is due to the motions of the different cores relative to one another, and this part 
is inherited by all stars; and secondly because the brown dwarfs are ejected with 
rather modest velocities ($\stackrel{<}{\sim} 1\,{\rm km}\,{\rm s}^{-1}$), and 
the higher-mass stars involved in the ejection also have increased velocity 
dispersion due to their recoil and their now being in a harder binary system. 
The main concern with these simulations is that, by invoking sink particles, 
protostellar embryos are instantaneously converted into point masses. This 
predisposes them to dynamical ejection, and prohibits them from merging or 
fragmenting further. Therefore the efficiency of the mechanism may have been 
overestimated.

Additional support for the mechanism comes from Goodwin et al. 
(2004b), who present an ensemble of simulations of the collapse and fragmentation 
of cores having a mass spectrum, density profiles, and low levels of turbulence, 
matched to those observed in Taurus. These simulations reproduce rather well the 
unusual stellar IMF observed in Taurus, including the 
relative paucity of brown dwarfs. As far as we are aware, these are the first 
simulations to demonstrate a direct causal link between the core mass spectrum and 
the stellar IMF.

Again we note that this mechanism does not exclude the previous two; indeed it 
requires formation by fragmentation of a collpasing core as a precursor to produce 
the low-mass protostellar embryos which then get ejected. However, ejection is 
unlikely to be the only mechanism forming brown dwarfs, since it seems very 
unlikely to produce the rather large numbers of close BD-BD binaries observed.

\section{Formation by photo-erosion} \label{SEC:EROS}

A fourth -- and somewhat separate -- mechanism for forming brown dwarfs is to 
start with a pre-existing core of standard mass (i.e. $\stackrel{>}{\sim} 
{\rm M}_{_\odot}$)and have it overrun by an 
HII region (Hester et al., 1996). As a result, an ionisation front (IF) starts to 
eat into the core, `photo-eroding' it. The IF is preceded by a compression wave 
(CW), and when the CW reaches the centre, a protostar is created, which then grows 
by accretion. At the same time, an expansion wave (EW) is reflected and propagates 
outwards, setting up the inflow which feeds accretion onto the central protostar. 
The outward propagating EW soon meets the inward propagating IF, and shortly 
thereafter the IF finds itself ionising gas which is so tightly bound to the 
protostar that it cannot be unbound by the act of ionisation. All the material 
interior to the IF at this juncture ends up in the protostar. On the basis of a 
simple semi-analytic treatment, Whitworth \& Zinnecker (2004) show that the final 
mass is given by
\begin{eqnarray} \nonumber
M & \sim & 0.01\,{\rm M}_{_\odot}\,
\left(\frac{a_{_{\rm I}}}{0.3\,{\rm km}\,{\rm s}^{-1}}\right)^6\, 
\left(\frac{\dot{\cal N}_{_{\rm LyC}}}{10^{50}\,{\rm s}^{-1}}\right)^{-1/3}\, \\
 & & \hspace{3.8cm} \times\,\left(\frac{n_{_{\rm O}}}{10^3\,{\rm cm}^{-3}}\right)^{-1/3},
\end{eqnarray}
where $a_{_{\rm I}}$ is the isothermal sound speed in the neutral gas of the core, 
$\dot{\cal N}_{_{\rm LyC}}$ is the rate at which the star(s) exciting the HII region 
emit hydrogen-ionising photons, and $n_{_{\rm O}}$ is the density in the ambient HII 
region.

This mechanism is rather robust, in the sense that it produces very low-mass stars for a 
wide range of initial conditions, and these conditions are likely to be realized in nature. 
Indeed, the evaporating gaseous globules (EGGs) identified in M16 by Hester et al. (1996) 
-- and subsequently in other HII regions -- would appear to be pre-existing cores 
being photo-eroded in the manner we have described. However, the mechanism is also 
very inefficient, in the sense that it usually takes a rather massive pre-existing 
prestellar core to form a single very low-mass star. Moreover, the mechanism can 
only work in the immediate vicinity of an OB star, so it cannot explain the 
formation of all brown dwarfs, and another mechanism is required to explain 
those seen in star formation regions like Taurus. Brown dwarfs formed in this way 
should include close BD-BD binaries, but they are unlikely to retain large accretion 
discs. 

\section{Conclusions} \label{SEC:CONC}

We have discussed four possible mechanisms for forming brown dwarfs: turbulent 
fragmentation, disc fragmentation, dynamical ejection and photo-erosion. None of 
these is mutually exclusive, and in fact the first three may occur consecutively. 
We emphasise that none of these mechanisms has been modelled properly with a 
fully radiative 3-D magneto-hydrodynamical code. Therefore, neither the thermal 
effects which presumably determine the minimum mass for star formation, nor the 
angular momentum transport processes which presumably determine the binary 
statistics, nor the $N$-body dynamics which presumably determine the clustering 
properties of brown dwarfs, has yet been properly evaluated.

\end{document}